\documentclass[12pt]{iopart}

\usepackage[final]{graphicx}
\usepackage{isotope}
\usepackage{citesort}

\usepackage{wignert}
\usepackage{mciop}
\newcommand{\Npair}{\mathcal{N}}

\newcommand{\MeV}{{\mathrm{MeV}}}

\def\C(#1){C_{#1}}
\def\Ct(#1){\tilde{C}_{#1}}
\def\Cd(#1){C_{#1}^\dagger}

\def\A(#1#2#3){A_{{#1}{#2}}^{#3}}
\def\At(#1#2#3){\tilde{A}_{{#1}{#2}}^{#3}}
\def\Ad(#1#2#3){A_{{#1}{#2}}^{#3\,\dagger}}

\def\Sdpow(#1){S^{\dagger\,#1}}
\def\Spow(#1){S^{#1}}
\def\sdpow(#1){s^{\dagger\,#1}}

\def\T(#1#2#3){T_{{#1}{#2}}^{#3}}
\def\TBOCCCC(#1#2#3#4#5#6#7){[(\Cd({#1})\times\Cd({#2}))^{#3}\times(\Ct({#4})\times\Ct({#5}))^{#6}]^{#7}}
\def\TBOAA(#1#2#3#4#5#6#7){(\Ad(#1#2#3)\times\At(#4#5#6))^{#7}}

\begin{document}
%bibliographystyle{iopart-num}

%***************************************************************************
% front matter
%***************************************************************************

\title{Generalized seniority with realistic interactions in open-shell nuclei}

\author{
M~A~Caprio$^1$,
F~Q~Luo$^1$,
K~Cai$^{1,2}$\footnote{Present address: Perimeter Institute for Theoretical Physics, 
Waterloo, Ontario N2L~2Y5, Canada.},
Ch~Constantinou$^1$ 
and~V~Hellemans$^{1,3}$
}
\address{$^1$~Department of Physics, University of Notre Dame,
Notre Dame, Indiana 46556-5670, USA}
\address{$^2$~Department of Physics, Bard College, 
Annandale-on-Hudson, New York 12504-5000, USA}
\address{$^3$~Physique Nucl\'eaire Th\'eorique et
Physique Math\'ematique, Universit\'e Libre de 
Bruxelles, CP229, B-1050 Brussels, Belgium}

%% \date{\today}

\begin{abstract}
Generalized seniority provides a truncation scheme for the nuclear
shell model, based on pairing correlations, which offers the
possibility of dramatically reducing the dimensionality of the nuclear
shell-model problem.  Systematic comparisons against results obtained
in the full shell-model space are required to assess the viability of
this scheme.  Here, we extend recent generalized seniority
calculations for semimagic nuclei, the $\isotope{Ca}$ isotopes, to
open-shell nuclei, with both valence protons and valence neutrons.
The even-mass $\isotope{Ti}$ and $\isotope{Cr}$ isotopes are treated
in a full major shell and with realistic interactions, in the
generalized seniority scheme with one broken proton pair and one
broken neutron pair.  Results for level energies, orbital occupations,
and electromagnetic observables are compared with those obtained in
the full shell-model space.  We demonstrate that, even for the $\isotope{Ti}$ isotopes, significant
benefit would be obtained in going beyond the approximation of one
broken pair of each type, while the $\isotope{Cr}$ isotopes require
further broken pairs to provide even qualitative accuracy.
\end{abstract}

\pacs{21.60.Cs,21.60.Ev}
% 21.10.Re (collective levels)
% 21.60.De ???
% 21.60.Cs (shell model), 
% 21.45.+v (few-body systems)
% 21.30.-x (nuclear forces)
% 21.60.Ev (collective models)

%% \maketitle

%***************************************************************************
% main text
%***************************************************************************

\section{Introduction}
\label{sec-intro}

The generalized seniority
scheme~\cite{talmi1971:shell-seniority,shlomo1972:gen-seniority}, or
broken pair approximation~\cite{gambhir1969:bpm,allaart1988:bpm}, has
long been proposed as a truncation scheme for the nuclear shell model,
based on pairing correlations, with the potential to dramatically
reduce the dimensionality of the nuclear shell-model problem.  The
generalized seniority scheme has also been proposed as a microscopic
foundation for the phenomenologically successful interacting boson
model (IBM)~\cite{iachello1987:ibm}, through the Otsuka-Arima-Iachello
mapping~\cite{otsuka1978:ibm2-shell-details,iachello1987:ibm-shell}.
The underlying premise of the generalized seniority scheme is that the
ground state of an even-even nucleus can be well approximated by a
condensate built from \textit{collective} $S$ pairs.  These are
defined as a specific linear combination of pairs of nucleons in the
different valence orbitals, each pair coupled to angular momentum
zero.  A shell-model calculation for the ground state and low-lying
states can then be carried out in a truncated space, consisting of
states built from a condensate of collective $S$ pairs together with a
small number $v$ (the
\textit{generalized seniority}) of nucleons not forming part of an $S$
pair.  

Although the generalized seniority approach has been applied in
various
contexts~\cite{gambhir1971:bpm-ni-sn,bonsignori1978:tda-sn,pittel1982:ibm-micro,scholten1983:gssm-n82,bonsignori1985:bpm-sn,vanisacker1986:ibm-cm-micro,navratil1988:ibfm-beta-a195-a197,engel1989:gssm-double-beta,lipas1990:ibm-micro-escatt,otsuka1996:ibm-micro,yoshinaga1996:ibm-micro-sm,monnoye2002:gssm-ni,barea2009:ibm-doublebeta},
only recently has it been benchmarked against calculations carried out
in the full shell-model space, with realistic interactions.
Comparisons have so far focused on semimagic
nuclei~\cite{sandulescu1997:gssm-sn,lei2010:pair-ca,caprio2012:gssmca}.
Systematic comparisons for both even-mass and odd-mass $\isotope{Ca}$
isotopes, across the full $pf$ shell ($20\leq N \leq 40$), with the
FPD6~\cite{richter1991:fpd6} and
GXPF1~\cite{honma2004:gxpf1-stability} interactions, are presented in~\cite{caprio2012:gssmca}.  These benchmark calculations for
semimagic
nuclei are based on the assumption of at most one broken pair,
\textit{i.e.}, $v=2$ for even-mass isotopes, which is found to
provide a quantitatively successful reproduction of many of the full-space
results (energies, occupations, and electromagnetic observables) for
the lowest-lying (and, in particular, yrast) states.  
In the interior of the shell, where both valence protons and neutrons
are present, the obvious challenge to the generalized seniority
truncation is the seniority-nonconserving, or pair-breaking, nature of
the proton-neutron quadrupole-quadrupole
interaction~\cite{talmi1983:ibm-shell}.  As this interaction induces
deformation, it also suppresses pairing correlations.  

The purpose of the present work is to investigate the viability of a
highly-truncated generalized seniority description as one moves beyond
semimagic nuclei, introducing valence nucleons of both types, and to
map the breakdown of this description.  
The generalized seniority scheme, as a simple representation of the
full nuclear shell-model problem, is foremost of conceptual interest.
The generalized seniority model space of generalized seniority $v$ is
equivalent to the space spanned by BCS states with at most $v$
quasiparticles, projected onto definite particle number.  Therefore,
the principal question being addressed by the present calculations is
to what extent the shell-model wave functions and predictions
quantitatively reflect a simple structure based on BCS-like pairing,
in a form which nonetheless fully accounts for particle number
conservation.  However, the generalized seniority scheme may also be
of computational relevance, as a practical truncation scheme.  In the
$pf$ shell, the dimensions of the full shell-model space for semimagic
and near-semimagic nuclei are computationally tractable.  However,
truncation is still necessary if single-particle spaces significantly
larger than the $pf$ shell are considered. A generalized seniority
truncation is more likely to be advantageous for weakly-deformed
nuclei near closed-shell, in large single-particle spaces, than for
strongly-deformed nuclei with large numbers of both protons and
neutrons in the valence shell.

We extend the
investigations of the $\isotope{Ca}$ isotopes
($Z=20$)~\cite{caprio2012:gssmca} into the interior of the $pf$ shell,
to the even-even $\isotope{Ti}$ ($Z=22$) and $\isotope{Cr}$ ($Z=24$)
isotopes, establishing a benchmark
comparison of results obtained in the generalized seniority truncation
against results obtained in the full shell-model space.  
We consider calculations truncated to one broken proton
pair plus one broken neutron pair,
\textit{i.e.}, $(v_p,v_n)=(2,2)$.
These calculations should be viewed as a baseline, in that they are
based on the most restricted generalized seniority truncation for both
protons and neutron.  Calculations involving more broken pairs could
be expected to provide a description of the dynamics which extends
further into the interior of the shell.  The present study is
therefore also intended to provide an indication of the dependences
upon valence proton and neutron numbers, interactions, and observables
likely to influence successful treatment of nuclei in such spaces of higher
generalized seniority.

The generalized seniority basis, in a
proton-neutron scheme, and computational method are summarized in section~\ref{sec-methods}.
Then, calculations for the $\isotope{Ti}$ and $\isotope{Cr}$ isotopes
in the generalized seniority scheme are compared with full shell-model
calculations, for level energies (section~\ref{sec-results-energy}),
orbital occupations (section~\ref{sec-results-occ}), and electromagnetic
observables (section~\ref{sec-results-trans}).  Preliminary results were
presented in~\cite{caprio2012:gssmpf-bpte12}.

\section{Generalized seniority scheme}
\label{sec-methods}

We briefly review the construction of the generalized seniority basis
for like
particles~\cite{pittel1982:ibm-micro,frank1982:ibm-commutator,allaart1988:bpm,caprio2012:gssmca}
and its extension to a proton-neutron
scheme.  
Let $\Cd(a,m_a)$ be the creation
operator for a particle in the shell-model orbital
$a\equiv(n_al_aj_a)$, with angular momentum projection quantum number
$m_a$. 
Then the collective $S$
pair of the generalized seniority scheme is defined by
$S^\dagger\equiv\sum_a \tfrac12 \alpha_a \hat{\jmath}_a
(\Cd(a)\Cd(a))^{(0)}$, where $a$ runs over the active orbitals, and $\hat{\jmath}_a\equiv (2j_a+1)^{1/2}$.  
This operator creates a linear combination of pairs in
different orbitals $a$, with amplitudes $\alpha_a$.

A basis state within the
generalized seniority scheme then consists of a ``condensate'' of
collective pairs, together with $v$ additional nucleons not
forming part of a collective $S$ pair.  The number $v$ is termed
the \textit{generalized seniority}.  Thus, if we consider
semimagic nuclei, so only like valence particles (all neutrons or all
protons) are present, the $v=0$ condensate state is
constructed as $\Sdpow(\Npair) \,\tket{}$, with angular momentum $J=0$. States with 
$v=2$ and angular momentum $J$ are then obtained as
$\Sdpow(\Npair-1)\,(\Cd(a)\Cd(b))^{(J)} \,\tket{}$, 
states with 
$v=4$ as
$\Sdpow(\Npair-2)\,[(\Cd(a)\Cd(b))^{(J_{12})}(\Cd(c)\Cd(d))^{(J_{34})}]^{(J)}
\,\tket{}$, \textit{etc.},
where, if $n$ is the number of valence nucleons, $\Npair=n/2$ is the total number of pairs.  

To construct a generalized seniority basis, the parameters $\alpha_a$
must first be determined.  This is usually accomplished variationally,
so as to minimize the energy functional
$E_\alpha=\tme{\Spow(\Npair)}{H}{\Spow(\Npair)}/\toverlap{\Spow(\Npair)}{\Spow(\Npair)}$~\cite{gambhir1969:bpm,otsuka1993:ibm2-ba-te-microscopic},
subject to the conventional normalization $\sum_a
(2j_a+1)\alpha_a^2=\sum_a(2j_a+1)$~\cite{pittel1982:ibm-micro}.  The
$S$-condensate state has the same form as a number-projected BCS
ground state, and these $\alpha_a$ coefficients are related to BCS
occupancy parameters $u_a$ and $v_a$ by
$\alpha_a=v_a/u_a$~\cite{allaart1988:bpm}.  The generalized seniority
states as defined above are in general unnormalized and nonorthogonal
(for $v\geq 3$, they also form an overcomplete set).  However, a
suitable basis is obtained by a Gram-Schmidt procedure, which yields
orthonormal basis states as linear combinations of the original
states,
\textit{e.g.},  for $v=2$, 
\begin{equation}
\tket{\Npair;v=2;J,k}=\sum_{ab} c_{ab;Jk}\tket{\Spow(\Npair-1)(\C(a)
\C(b))^{(J)}},
\end{equation}
where $k$ is simply a counting index for the orthogonalized states.
Several
approaches~\cite{frank1982:ibm-commutator,vanisacker1986:ibm-cm-micro,allaart1988:bpm,mizusaki1996:ibm-micro,chen1997:npsm}
have been developed for evaluating matrix elements of one-body and two-body operators in the generalized
seniority basis.  The present calculations
make use of recurrence relations derived in~\cite{luo2011:gssmme}, where the notation and methods used in
the present work are also established in detail.
%----------------------------------------------------------------
\begin{figure}
\begin{center}
\includegraphics*[width=0.5\hsize]{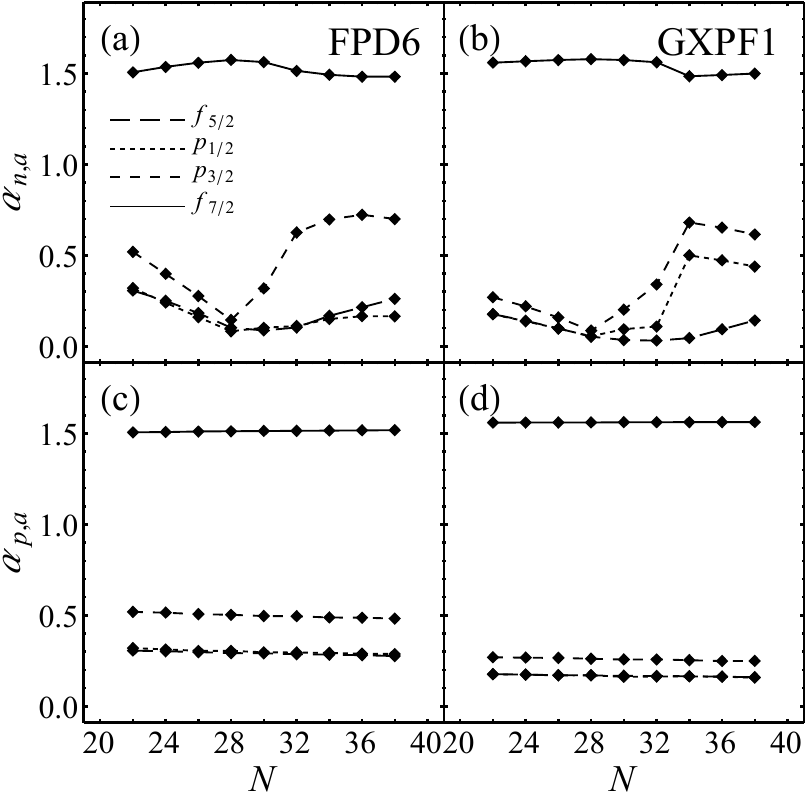}
\end{center}
\caption{Structure of the collective $S$ pairs for $\isotope{Ti}$, for neutrons~(top) and protons~(bottom), and for
the FPD6~(left) and GXPF1~(right) interactions.  Amplitudes are obtained by the variational prescription (see text).
}
\label{fig-alpha}
\end{figure}
%----------------------------------------------------------------

A generalized seniority basis can be defined for nuclei with valence
particles of both types via a proton-neutron scheme, that is, by
taking all possible products of proton and neutron generalized
seniority states, with generalized seniorities $v_p$ and
$v_n$~\cite{allaart1988:bpm,monnoye2002:gssm-ni}.  Here we consider states with one
broken pair of each type [$(v_p,v_n)=(2,2)$] and thus have
basis states
\begin{equation}
\bigl[\tket{\Npair_p;v_p=2;J_p,k_p}\otimes\tket{\Npair_n;v_n=2;J_n,k_n}\bigr]^{(J)}.
\end{equation}
The coefficients $\alpha_{p,a}$ appearing in the proton collective
pair operator [$S^\dagger_p\equiv\sum_a \tfrac12 \alpha_{p,a}
\hat{\jmath}_a (\Cd(p,a)\Cd(p,a))^{(0)}$] and the coefficients
$\alpha_{n,a}$ appearing in the neutron collective pair operator
[$S^\dagger_n\equiv\sum_a \tfrac12 \alpha_{n,a} \hat{\jmath}_a
(\Cd(n,a)\Cd(n,a))^{(0)}$] are distinct, and these are chosen
independently, by minimizing
$E_{p,\alpha_p}=\tme{\Spow(\Npair_p)_p}{H}{\Spow(\Npair_p)_p}/\toverlap{\Spow(\Npair_p)_p}{\Spow(\Npair_p)_p}$
and
$E_{n,\alpha_n}=\tme{\Spow(\Npair_n)_n}{H}{\Spow(\Npair_n)_n}/\toverlap{\Spow(\Npair_n)_n}{\Spow(\Npair_n)_n}$,
respectively.  The values of the amplitudes $\alpha_{p,a}$ and
$\alpha_{n,a}$ obtained in this manner for the $\isotope{Ti}$ isotopes
are shown in figure~\ref{fig-alpha}, for both the FPD6 and GXPF1
interactions.  The variational expression $E_{p,\alpha_p}$ for the
proton $S$ pair has no manifest dependence upon neutron number and,
similarly, the variational expression $E_{n,\alpha_n}$ for the neutron
$S$ pair has no manifest dependence upon proton number.  However, an
implicit dependence is induced, in each case, by the $A$-dependent
definitions of the interaction two-body matrix elements, which are
proportional to $A^{-0.35}$ for FPD6~\cite{richter1991:fpd6} or
$A^{-0.3}$ for GXPF1~\cite{honma2004:gxpf1-stability}.  A simple
multiplicative scaling of the Hamiltonian would not affect the collective pair
structure, but note that here rather the $A$-dependent two-body contribution varies in strength relative to the $A$-independent
single-particle energies.  Consequently, the values of
$\alpha_{n,a}$ in the shell interior vary
slightly ($\lesssim 1 \%$) between successive isotopic chains, and the
values of $\alpha_{p,a}$ vary slightly along each isotopic chain
[figure~\ref{fig-alpha} (bottom)].

\begin{figure}
\begin{center}
\includegraphics*[width=\hsize]{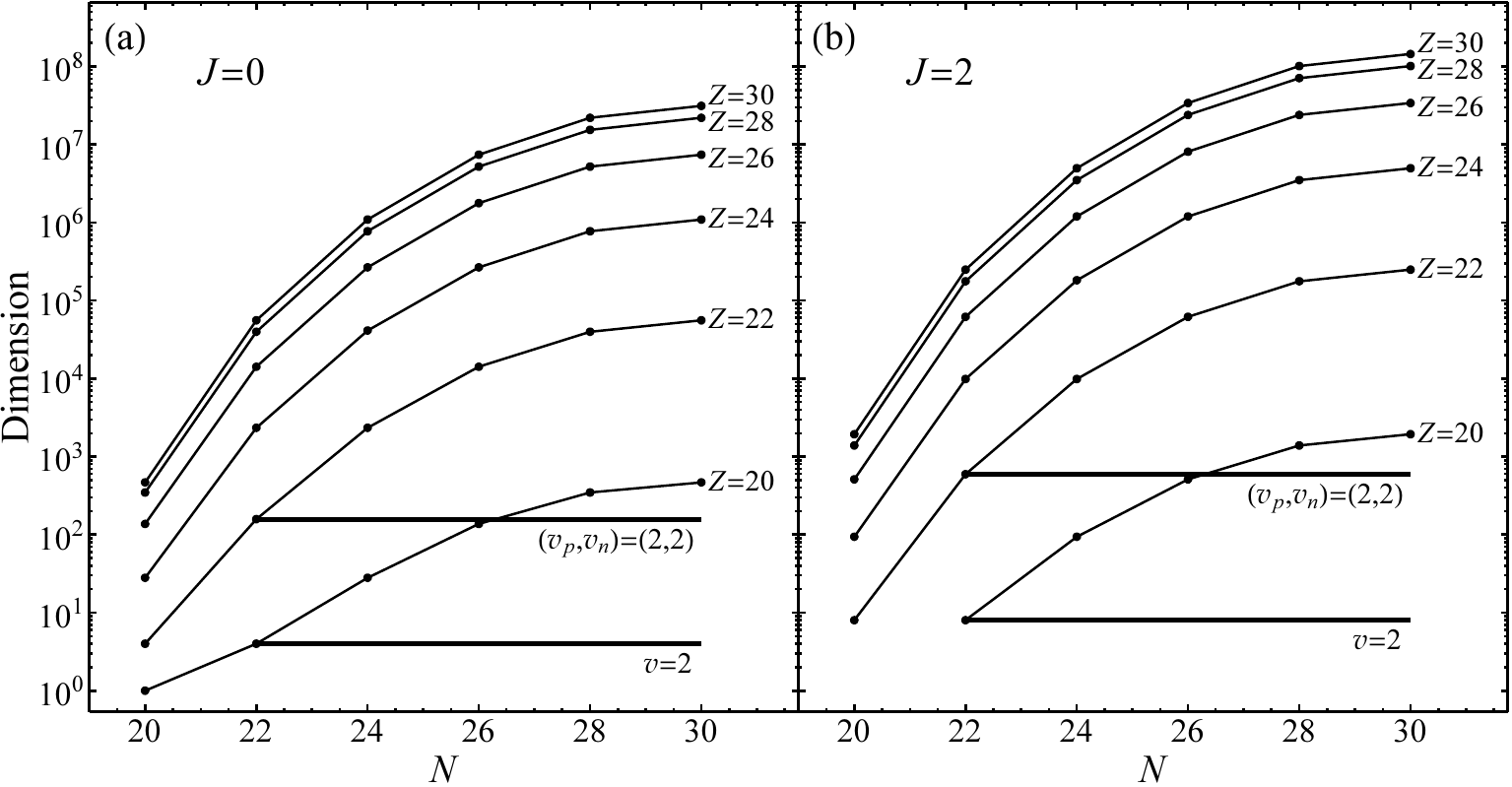}
\end{center}
\caption{
Dimensions of the $J$-scheme shell-model spaces for nuclei in the $pf$
shell, with $20\leq N,Z \leq 30$, shown for
(a)~$J=0$ and (b)~$J=2$.  The dimension of the generalized seniority
model space with one broken pair ($v=2$) is shown for
comparison with the semimagic $\isotope{Ca}$ isotopes, and the dimension of the generalized seniority
model space with one
broken pair of each type [$(v_p,v_n)=(2,2)$] is shown for
comparison with open-shell nuclei.  Dimensions
for nuclei with $30< N,Z \leq 40$ follow by particle-hole conjugation symmetry.
}
\label{fig-dim-pf}
\end{figure}

For semimagic nuclei, the size of the generalized seniority basis is
the same as for the shell-model problem with only $v$ particles in the
same set of orbitals, regardless of the number of $S$ pairs.  Thus,
the generalized seniority model space with one broken pair ($v=2)$ for any even
$\isotope{Ca}$ isotope taken in the $pf$ shell has the same dimension
as the shell-model space for $\isotope[42]{Ca}$~--- dimension $4$ for
$J=0$, $8$ for $J=2$, $6$ for $J=4$,
\textit{etc.}~\cite{caprio2012:gssmca}.  Similarly, if both valence
protons and neutrons are present, the basis size is the same as for
the shell-model problem with only $v_p$ protons and $v_n$ nucleons.
Thus, the generalized seniority model space with one broken proton
pair and one broken neutron pair [$(v_p,v_n)=(2,2)$] for any
even-even nucleus taken in the $pf$ shell has the same dimension as
the shell-model space for $\isotope[44]{Ti}$~--- dimension $158$ for
$J=0$, $596$ for $J=2$, $655$ for $J=4$, \textit{etc.}  For
comparison, the dimensions of the full shell-model spaces in the $pf$
shell are shown, in the $J$ scheme, in figure~\ref{fig-dim-pf}.
%----------------------------------------------------------------
%----------------------------------------------------------------
The
reduction in model space dimension from the full shell-model space to
the generalized seniority truncated space considered in this work
exceeds two orders of magnitude for the $\isotope{Ti}$ isotopes
midshell and approaches four orders of magnitude for the
$\isotope{Cr}$ isotopes midshell (\textit{e.g.}, factors of
$\sim4\times10^2$ and $\sim7\times10^3$ in the $J=0$ spaces for
$\isotope[52]{Ti}$ and $\isotope[54]{Cr}$, respectively).

\section{Results}
\label{sec-results}

\subsection{Energies}
\label{sec-results-energy}

Here we consider some basic energy observables, for the
even-even $\isotope{Ti}$ ($Z=22$) and $\isotope{Cr}$ ($Z=24$)
isotopes, in the generalized seniority $(v_p,v_n)=(2,2)$
space.  Calculations are shown for $20<N<40$, carried
out in the full $pf$ shell ($f_{7/2}$, $p_{3/2}$, $p_{1/2}$, and
$f_{5/2}$ orbitals) for both protons and neutrons, with the
FPD6~\cite{richter1991:fpd6} and
GXPF1~\cite{honma2004:gxpf1-stability} interactions.  
The generalized seniority results
are benchmarked against those obtained in the full shell-model space,
calculated using the code \textsc{NuShellX}~\cite{rae:nushellx-COMBO}.

\begin{figure}
\begin{center}
\includegraphics*[width=\hsize]{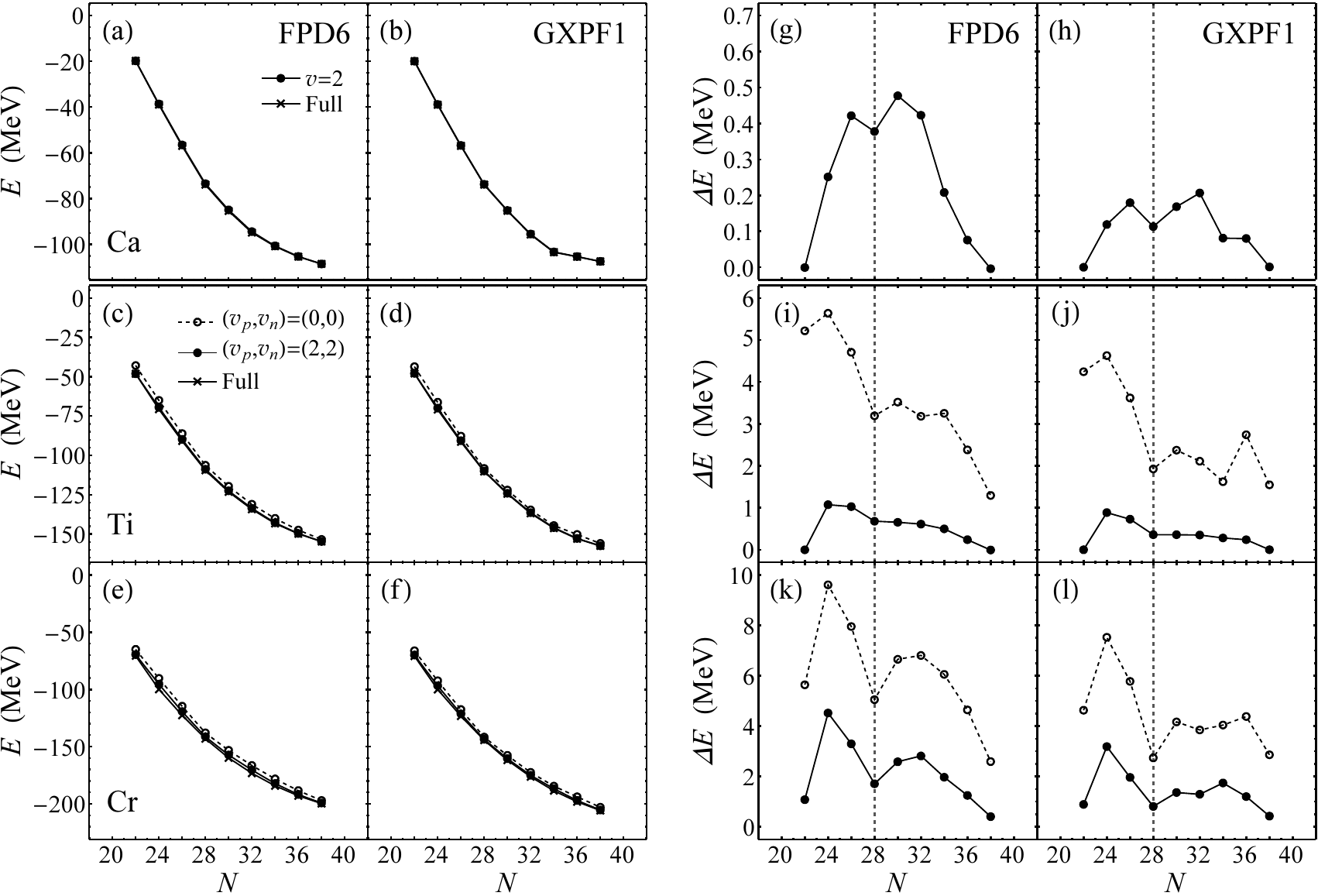}
\end{center}
\caption{
(Left)~Energy eigenvalue of the $0^+$ ground state, calculated in the
generalized seniority $S$-pair condensate (open circles),
$(v_p,v_n)=(2,2)$ model space (or $v=2$ for $\isotope{Ca}$) (solid circles) and full shell-model space (crosses).
(Right)~Residual energy difference $\Delta E$ of the generalized
seniority result relative to the full shell-model result.  Values are
shown for the even-mass $\isotope{Ca}$, $\isotope{Ti}$, and
$\isotope{Cr}$ isotopes (top to bottom, respectively), for the FPD6
and GXPF1 interactions.  The dashed vertical line indicates the $N=28$
subshell closure. Calculations for $\isotope{Ca}$ are from~\cite{caprio2012:gssmca}.  }
\label{fig-gs}
\end{figure}

We begin by considering the energy eigenvalue for the $J=0$ ground
state.  The calculated values are summarized in figure~\ref{fig-gs}
(left).  The ground
state energy is shown as obtained both for the $(v_p,v_n)=(0,0)$
proton-neutron $S$-pair condensate and in the $(v_p,v_n)=(2,2)$
generalized seniority model space, as well as in the full shell-model space.  
%----------------------------------------------------------------
%----------------------------------------------------------------

To provide a baseline for comparison, it is helpful to review the
analogous results for the semimagic $\isotope{Ca}$ isotopes in the
$v=2$ space, from~\cite{caprio2012:gssmca}, shown at top in
figure~\ref{fig-gs}.  It should first be noted (see appendix of~\cite{caprio2012:gssmca}) that the ground state obtained in the
$v=2$ space is simply the $v=0$ $S$-pair condensate, provided the $\alpha_a$
coefficients have been chosen variationally as in
section~\ref{sec-methods}.  Therefore, the ground-state energies obtained
in $v=0$ and $v=2$ spaces are identical.  [In contrast, for nuclei in
the interior of the shell, the ground state obtained in the
$(v_p,v_n)=(2,2)$ space is not in general the $(v_p,v_n)=(0,0)$
$S$-pair condensate, and these states are considered separately in
figure~\ref{fig-gs}.]  The ground state energies
[figure~\ref{fig-gs}(a,b)] obtained for $\isotope{Ca}$ in the
generalized seniority truncation are essentially indistinguishable
from those obtained in the full shell-model space, when viewed on the
$\sim100\,\MeV$ scale of the valence shell interaction energies.  A
more useful measure of the level of agreement is provided by the
residual energy difference $\Delta E$, shown in figure~\ref{fig-gs}
(right), obtained by subtracting the full space result from the
generalized seniority result.  This difference may be thought of as
the missing correlation energy, not accounted for by the generalized
seniority description of the ground state.  For the $\isotope{Ca}$
isotopes [figure~\ref{fig-gs}(g,h)], the maximum difference occurs
approximately midshell, peaking at $0.48\,\MeV$ for the FPD6
interaction or $0.21\,\MeV$ for the GXPF1 interaction.  It is worth
noting at this point that the deviations between the generalized
seniority results and the full-space results were found to be consistently smaller
for the GXPF1 interaction than for the FPD6 interaction in the study
of $\isotope{Ca}$ isotopes~--- not just for the ground state energy,
but for excitation energies, orbital occupations
(section~\ref{sec-results-occ}), and electromagnetic observables
(section~\ref{sec-results-trans}), as well.

Considering now the ground-state eigenvalues for $\isotope{Ti}$ in the
generalized seniority scheme [figure~\ref{fig-gs}(c,d)], a
$(v_p,v_n)=(0,0)$ calculation, which simply takes the expectation
value of the Hamiltonian in the proton-neutron $S$ pair condensate,
misses $\sim2$--$6\,\MeV$ in binding energy.  However, the
$(v_p,v_n)=(2,2)$ model space accounts for most of the missing
correlation energy, leaving a difference of $\lesssim1\,\MeV$ between
the generalized seniority and full-space results
[figure~\ref{fig-gs}(i,j)].  Since $\isotope{Ti}$ has two valence
protons in the $pf$ shell, a $(v_p,v_n)=(2,2)$ calculation encompasses
the full space of proton configurations.  Moreover, for $N=22$
($\isotope[44]{Ti}$) and $N=38$ ($\isotope[60]{Ti}$), there are only
two valence neutrons or neutron holes, so the $(v_p,v_n)=(2,2)$ space
is in fact identical to the full shell-model space, and the
generalized seniority and full shell-model results are strictly
identical.  The generalized seniority results lie furthest from the
full-space results in the middle of the $f_{7/2}$ subshell ($20< N <
28$), with a maximum deviation of $1.1\,\MeV$ for $N=24$
($\isotope[46]{Ti}$) with the FPD6 interaction or $0.9\,\MeV$ with the
GXPF1 interaction.  The generalized seniority results more closely
match the full-space results in the upper $pf$ shell ($28 < N < 40$),
where the average deviations are $0.47\,\MeV$ for
the FPD6 interaction or $0.28\,\MeV$ for the GXPF1 interaction.

For the evolution of $\isotope{Cr}$ ground-state eigenvalues across
the shell [figure~\ref{fig-gs}(e,f)], similar qualitative observations
apply as to the $\isotope{Ti}$ isotopes.  However, the energy
differences between the generalized seniority and full shell-model
results are much larger, up to $\sim 10\,\MeV$ for the
$(v_p,v_n)=(0,0)$ calculation.  Breaking one pair of each type yields
a less dramatic improvement than observed for the $\isotope{Ti}$
isotopes.  The generalized seniority results $(v_p,v_n)=(2,2)$ again
lie furthest from the full-space results in the middle of the
$f_{7/2}$ subshell, missing $4.5\,\MeV$ of binding energy for
$N=24$ ($\isotope[48]{Cr}$) with FPD6 or $3.2\,\MeV$ with GXPF1.  The
generalized seniority results again more closely match the full-space
results in the upper $pf$ shell, where the average deviations are
$2.0\,\MeV$ for the FPD6 interaction or $1.3\,\MeV$ for the GXPF1
interaction.  It has been hypothesized by Monnoye \textit{et
al.}~\cite{monnoye2002:gssm-ni} that the generalized seniority
description should improve at subshell closures.  The residual energy
difference does indeed have a sharp local minimum at $N=28$, for both
interactions, in both the $S$-condensate [$(v_p,v_n)=(0,0)$] and
broken-pair [$(v_p,v_n)=(2,2)$] calculations.
%----------------------------------------------------------------
\begin{table}
  \caption{Average residual energy differences $\Delta E$ (in $\mathrm{MeV}$)
  of the energy eigenvalues calculated in the generalized seniority
  $(v_p,v_n)=(2,2)$ model space (or $v=2$ for $\isotope{Ca}$) relative
  to the full shell-model results, for selected
  states of the $\isotope{Ca}$, $\isotope{Ti}$, and
  $\isotope{Cr}$ isotopes, and for the FPD6 and GXPF1 interactions.  These
  are root-mean-square averages over the full set of even-mass
  isotopes with $20< N<40$.  Calculations for $\isotope{Ca}$ are
  from~\cite{caprio2012:gssmca}.  }
\label{tab-energy}
\begin{center}
\begin{ruledtabular}
%%\begin{tabular}{ldddddddd} % iopart does not support d alignment even with dcolumn loaded
% note: use iopart \centre, \ns, and \crule -- regular \cline doesn't leave horizontal gap where needed
\begin{tabular}{lrrrrrrrr}
\br
&\centre{4}{FPD6}&\centre{4}{GXPF1}\\ \ns\ns
&\crule{4}&\crule{4}\\ 
&
\multicolumn{1}{c}{$0^+_1$}&\multicolumn{1}{c}{$2^+_1$}&\multicolumn{1}{c}{$4^+_1$}&\multicolumn{1}{c}{$0^+_2$}&
\multicolumn{1}{c}{$0^+_1$}&\multicolumn{1}{c}{$2^+_1$}&\multicolumn{1}{c}{$4^+_1$}&\multicolumn{1}{c}{$0^+_2$}\\
\mr
\isotope{Ca}&
0.31&0.48&0.62&0.88&    
0.13&0.25&0.34&0.63\\
\isotope{Ti}&
0.65&1.10&1.14&2.68&
0.45&0.76&0.82&1.66\\
\isotope{Cr}&
2.48&4.54&5.07&5.74&
1.62&3.04&3.55&4.02\\
\br
\end{tabular}
\end{ruledtabular}
\raggedright
\end{center}
\end{table}
%----------------------------------------------------------------

In considering excitation energies, we focus on the low-lying states,
for which the generalized seniority description was found to be most
successful in the semimagic nuclei, especially the first $2^+$ and
$4^+$ states.  The succussful reproduction of the properties of the
first excited $0^+$ state requires a greater number of broken pairs,
especially in the $f_{7/2}$ subshell~\cite{caprio2012:gssmca}.  For
the energy eigenvalues of the ground state and other low-lying states,
the average (root-mean-square) deviations of the generalized seniority
$(v_p,v_n)=(2,2)$ model space results from the full shell-model
results across the shell are summarized in table~\ref{tab-energy},
along with the corresponding $v=2$ results for $\isotope{Ca}$.

\begin{figure}
\begin{center}
\includegraphics*[width=\hsize]{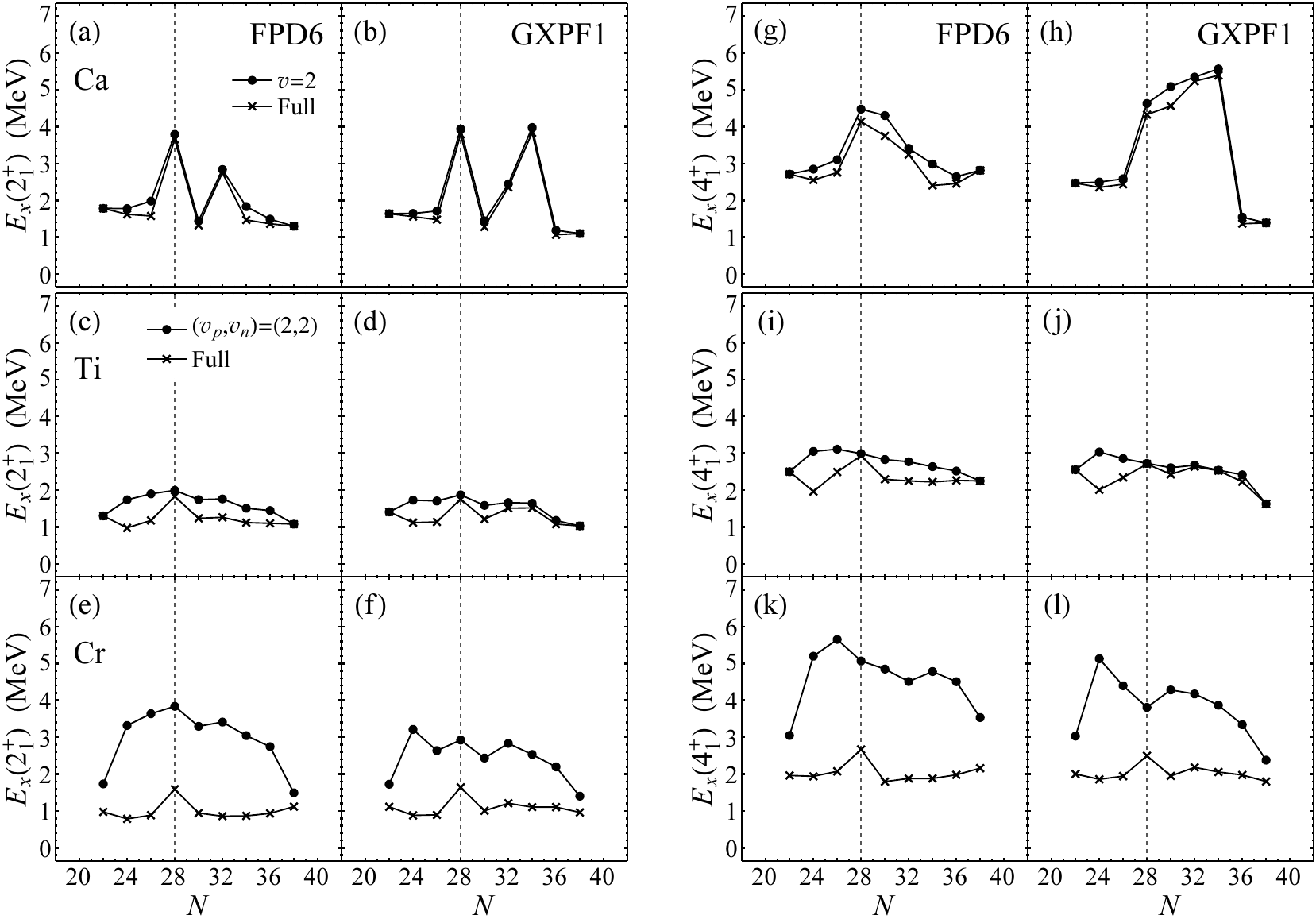}
\end{center}
\caption{
Excitation energies $E_x$ of the first $2^+$~(left) and $4^+$~(right) states, calculated in the generalized seniority
$(v_p,v_n)=(2,2)$ model space (or $v=2$ for $\isotope{Ca}$) (circles) and full shell-model space (crosses).
Values are shown for the even-mass $\isotope{Ca}$, $\isotope{Ti}$, and $\isotope{Cr}$
isotopes (top to bottom, respectively), for the FPD6 and GXPF1 interactions.
The dashed vertical line indicates the $N=28$
subshell closure. Calculations for $\isotope{Ca}$ are from~\cite{caprio2012:gssmca}.  
}
\label{fig-ex}
\end{figure}

The excitation energies $E_x$ of the first $2^+$ and $4^+$ states,
calculated relative to the $0^+$ ground state, are shown in
figure~\ref{fig-ex}.  We again begin by reviewing the results for the
$\isotope{Ca}$ isotopes, at top in figure~\ref{fig-ex}.  The broad
features of the evolution of excitation energies across the shell are
reproduced within the $v=2$ model space.  For instance, for the $2^+$
state [figure~\ref{fig-ex}(a,b)], spikes in excitation energy are
obtained at the $f_{7/2}$ subshell closure ($N=28$) and $p_{3/2}$
subshell closure ($N\approx32$--$34$).  The excitation energy
calculated for the $2^+$ state deviates from that calculated in the
full model space by at most $0.41\,\MeV$ for FPD6 or $0.23\,\MeV$ for
GXPF1.  The deviations in excitation energy for the $4^+$ state
[figure~\ref{fig-ex}(g,h)] are only modestly larger, at most
$0.58\,\MeV$ for FPD6 or $0.53\,\MeV$ for GXPF1.
%----------------------------------------------------------------
%----------------------------------------------------------------

For the excitation energies of the $2^+$ [figure~\ref{fig-ex}(c,d)] and
$4^+$ [figure~\ref{fig-ex}(i,j)] states in the $\isotope{Ti}$ isotopes,
the values obtained in the generalized seniority truncated space and
the full shell-model space necessarily agree at $N=22$ and $N=38$,
where, as noted above, the $(v_p,v_n)=(2,2)$ space is identical to the
full space.  However, otherwise, substantial differences arise.  It is
interesting to note the cases in which these calculations at low
generalized seniority are most successful.  Differences are largest in
the $f_{7/2}$ subshell.  Then, at the subshell closure ($N=28$), the
excitation energies calculated in the generalized seniority truncated
space closely reproduce those in the full space~--- moderately closely
for the $2^+$ energy, to $0.17\,\MeV$ for FPD6 and $0.12\,\MeV$ for
GXPF1, and much more closely for the $4^+$ energy, to $0.060\,\MeV$
(or $\sim2\%$) for FPD6 and $0.017\,\MeV$ (or $\sim0.6\%$) for GXPF1.
In the upper $pf$ shell, specifically for the GXPF1 interaction, the excitation
energies of the $2^+$ [figure~\ref{fig-ex}(d)] and $4^+$ states
[figure~\ref{fig-ex}(j)] are reproduced with a quantitative accuracy
comparable to that observed in the semimagic nuclei.  Deviations
from the full shell-model results average only $0.19\,\MeV$
for the $2^+$ state or $0.12\,\MeV$ for the
$4^+$ state.  The excitation energies obtained in the generalized
seniority truncated model space are systematically higher than those
in the full model space, \textit{i.e.}, more correlation energy is
missed in the excited state eigenvalue than in the ground state
eigenvalue.

For the $\isotope{Cr}$ isotopes, it is seen that considering only one broken pair of
each type is markedly inadequate for description of the excitation
energies, for both the $2^+$ [figure~\ref{fig-ex}(e,f)] and $4^+$
[figure~\ref{fig-ex}(k,l)] states.  The actual excitation energies calculated in the full space
are comparable to those for the $\isotope{Ti}$ isotopes, but the
excitation energies obtained in the generalized seniority truncation
are about twice as high.

\subsection{Occupations}
\label{sec-results-occ}

\begin{figure}
\begin{center}
\includegraphics*[width=\hsize]{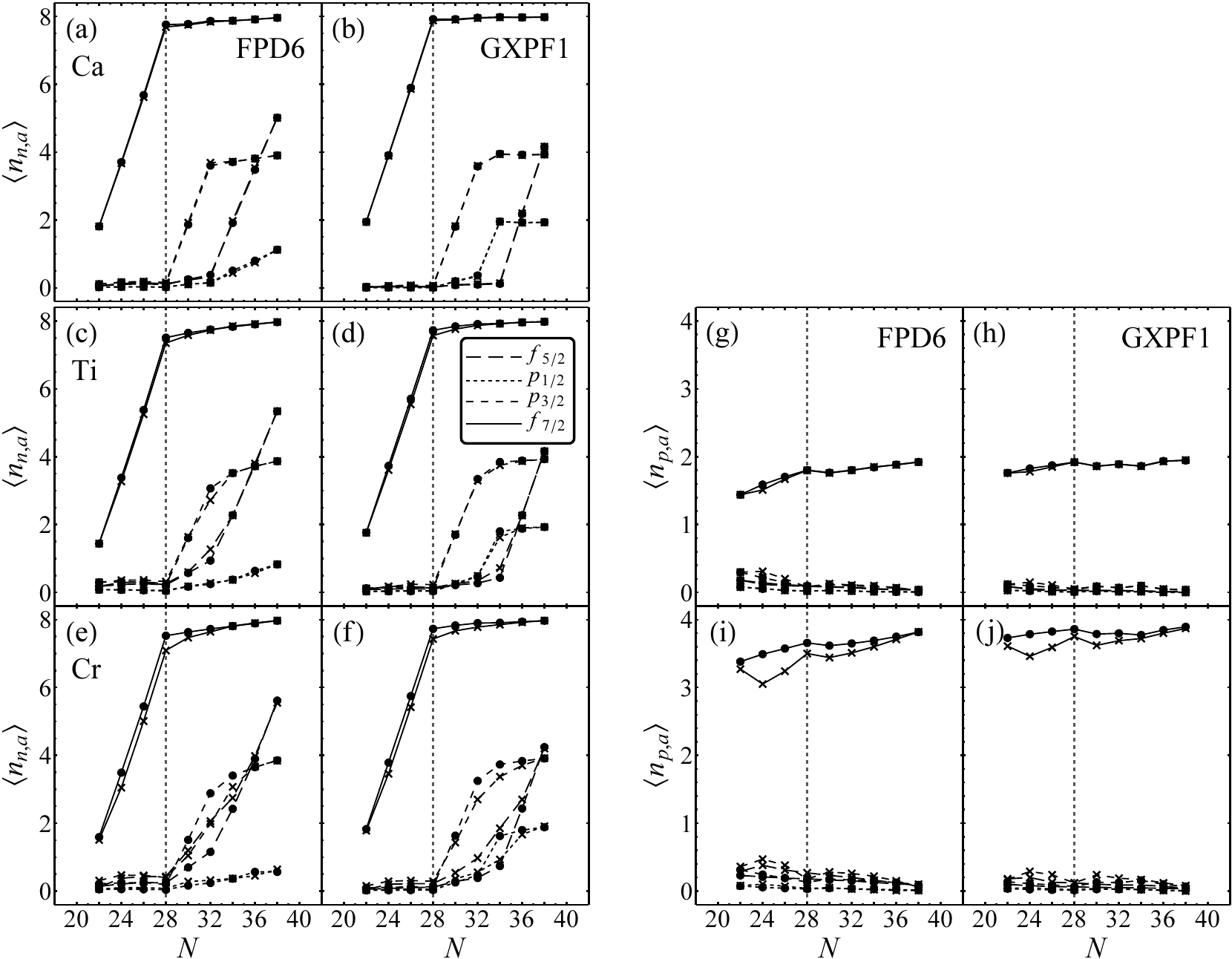}
\end{center}
\caption{Orbital occupations for
the $0^+$ ground state, for neutron (left) and proton (right) orbitals, calculated in the generalized seniority
$(v_p,v_n)=(2,2)$ model space (or $v=2$ for $\isotope{Ca}$) (circles) and full shell-model space (crosses).
Values are shown for the even-mass $\isotope{Ca}$, $\isotope{Ti}$, and $\isotope{Cr}$
isotopes (top to bottom, respectively), for the FPD6 and GXPF1 interactions.
The dashed vertical line indicates the $N=28$
subshell closure. Calculations for $\isotope{Ca}$ are from~\cite{caprio2012:gssmca}.  }
\label{fig-occ}
\end{figure}

The occupations of orbitals provide a direct measure of the structure
of an eigenstate, here as obtained either in a generalized seniority
calculation or the full shell-model space.  Unlike conventional
shell-model basis states, the generalized seniority basis states do
not have definite occupation for each orbital, rather, involving a
BCS-like distribution of occupations.  Therefore, the occupation
$\tbracket{n_a}$ of an orbital $a=(n_al_aj_a)$ in an eigenstate represented
in this basis cannot be obtained as a simple average over the
contributing basis states, but rather must be evaluated as the
expectation value of a one-body operator, the number operator for the
orbital [$n_a=-\hat{\jmath}_a (\Cd(a)\Ct(a))^{(0)}$], by the process
described in~\cite{luo2011:gssmme,caprio2012:gssmca}.  Occupations of
each of the $pf$-shell orbitals in the $0^+$ ground state are shown in
figure~\ref{fig-occ}, for the $\isotope{Ca}$, $\isotope{Ti}$, and
$\isotope{Cr}$ isotopes, both for neutron orbitals
[figure~\ref{fig-occ} (left)] and proton orbitals
[figure~\ref{fig-occ} (right)].
%----------------------------------------------------------------
%----------------------------------------------------------------
%----------------------------------------------------------------
\begin{figure}
\begin{center}
\includegraphics*[width=\hsize]{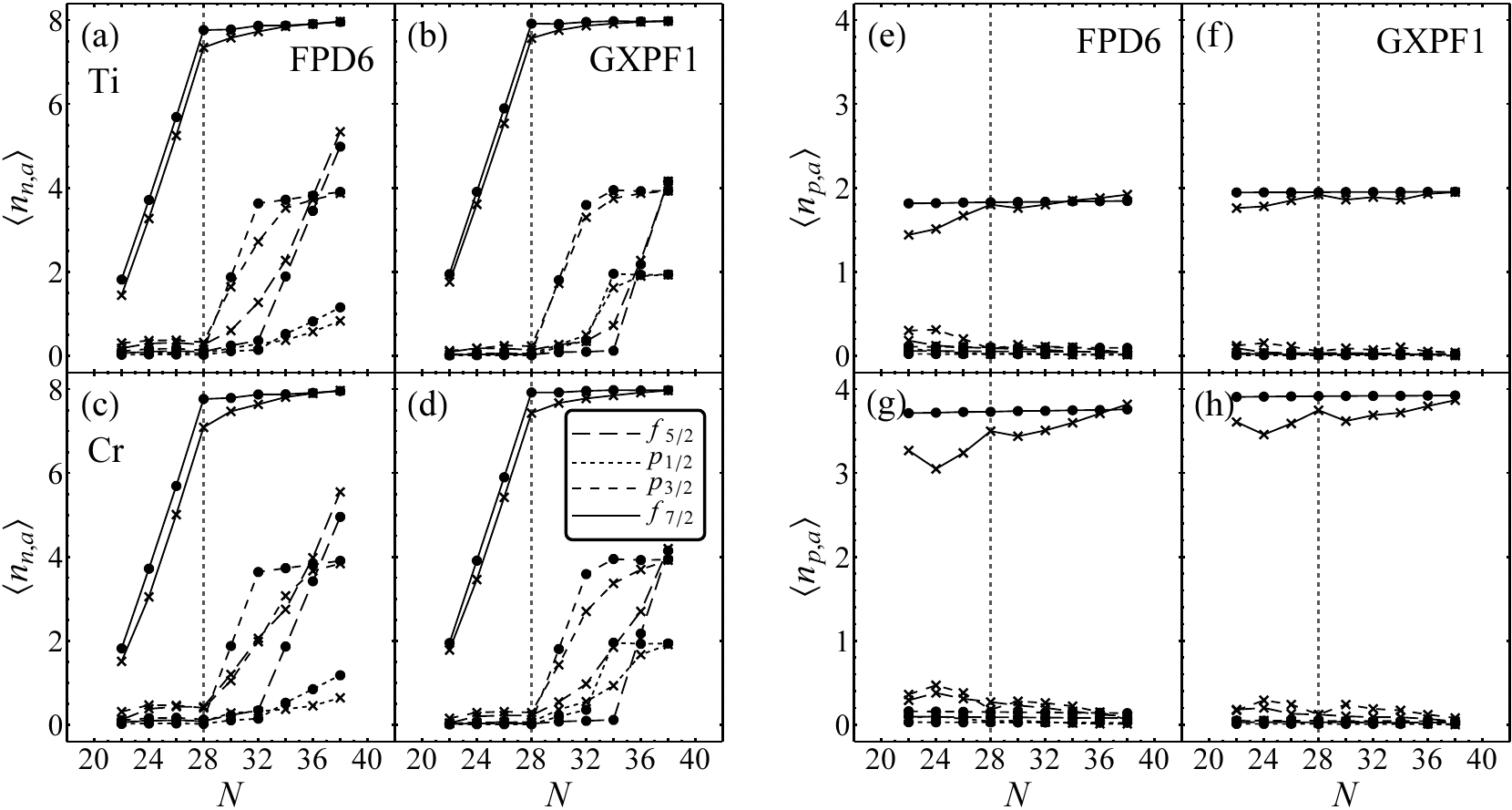}
\end{center}
\caption{Orbital
occupations for
the $0^+$ ground state described as the $(v_p,v_n)=(0,0)$ proton-neutron $S$-pair
condensate (circles)~--- for comparison with those obtained 
in the $(v_p,v_n)=(2,2)$ model space in
figure~\ref{fig-occ}~--- for the even $\isotope{Ti}$ and
$\isotope{Cr}$ isotopes (top to bottom, respectively), for neutron
(left) and proton (right) orbitals, and for the FPD6 and GXPF1
interactions.  Occupations obtained in the full shell-model space
(crosses) are shown again as well.  The dashed vertical line indicates
the $N=28$ subshell closure. }
\label{fig-occ-v0}
\end{figure}
%----------------------------------------------------------------

We again begin by reviewing the situation for the $\isotope{Ca}$
isotopes [figure~\ref{fig-occ}(a,b)].  Recall that the ground state
obtained in the $v=2$ space, for semimagic nuclei, is simply the $v=0$
$S$-pair condensate (section~\ref{sec-results-energy}).  For $20\leq N
\leq 28$, the neutrons are almost exclusively found in the $f_{7/2}$
orbital.  Therefore, the generalized seniority scheme essentially
reduces to conventional seniority in a single $j$-shell.  However, in
the upper $pf$ shell, for $28< N < 40$, neutrons fill multiple
orbitals simultaneously, and the generalized seniority scheme
prescribes correlations among these $j$-shells beyond the conventional
seniority correlations within each $j$-shell.  The generalized
seniority calculations closely reproduce the occupations obtained in
the full shell-model space, to within $0.1$ nucleon across the entire
shell, and with deviations averaging only $0.03$ nucleon for FPD6 or
$0.015$ for GXPF1.  The occupations obtained in the first $2^+$ and
$4^+$ states follow very similar patterns to those for the ground
state.  For
the first excited $0^+$ state, the occupations are only well-described
in the upper $pf$ shell (see figure~5 of~\cite{caprio2012:gssmca})~---
for $N<28$, the single-particle energies strongly favor an excitation
involving promotion of a single particle from the $f_{7/2}$ orbital to
the $p_{3/2}$ orbital, which is not supported in the highly-restricted
(dimension $4$)
$J=0$, $v=2$ space.

Let us consider now the occupations of these same neutron orbitals for
the $0^+$ ground state of the $\isotope{Ti}$ isotopes
[figure~\ref{fig-occ}(c,d)], as obtained in the $(v_p,v_n)=(2,2)$
generalized seniority space.  The occupations are reproduced to within
$0.3$ nucleon throughout the shell, with deviations averaging $0.10$
nucleon for FPD6 or $0.08$ for GXPF1~--- greater than for the
semimagic nuclei, but still leaving the generalized seniority and
full-space curves in figure~\ref{fig-occ}(c,d) largely
indistinguishable.  The deviations which do arise are localized in $N$
and are seen to follow a systematic pattern.  Over most of the shell,
but especially around the $f_{7/2}$ subshell closure at $N=28$, the
generalized seniority calculation overestimates the occupation of the
neutron $f_{7/2}$ orbital.  Similarly, at the nominal $p_{3/2}$
subshell closure at $N=32$, the generalized seniority calculations
overestimate the $p_{3/2}$ occupation relative to the $f_{5/2}$
occupation.  In this limited set of examples, the generalized
seniority calculations would appear to favor a filling order more
strictly reflecting the ordering of single-particle energies
($f_{7/2}$, $p_{3/2}$, $p_{1/2}$, $f_{5/2}$) than is actually found in
the solutions in the full shell-model space.  It is worth noting that
the quantitative accuracy obtained for the $\isotope{Ti}$ occupations
in figure~\ref{fig-occ}(c,d) is obtained only through the breaking of
collective $S$ pairs.  For comparison, occupations calculated in the
$(v_p,v_n)=(0,0)$ proton-neutron $S$-pair condensate for the
$\isotope{Ti}$ isotopes are shown in figure~\ref{fig-occ-v0}~(a,b).
These do reflect the overall evolution of the neutron occupations with
$N$, but with substantially larger deviations, averaging $0.31$ nucleon for FPD6 or
$0.20$ nucleon for GXPF1.

For the $\isotope{Cr}$ isotopes [figure~\ref{fig-occ}(e,f)], the
accuracy with which the neutron orbital occupations
for the $0^+$ ground state are reproduced within the generalized
seniority truncation deteriorates, averaging $0.30$ nucleon for both FPD6 and
GXPF1.  The systematic deviations continue to follow the trend
described above for the $\isotope{Ti}$ isotopes~--- in which the
generalized seniority results follow nominal filling order more
strictly than the full shell-model results~--- but now
increased in magnitude, extending over larger ranges of neutron
number, and observed in an overestimate of the $p_{1/2}$ orbital
occupation relative to $f_{5/2}$ as well.
%----------------------------------------------------------------
\begin{table}
  \caption{Average deviations of neutron orbital occupations $\tbracket{n_{n,a}}$
  calculated in the generalized seniority
  $(v_p,v_n)=(2,2)$ model space (or $v=2$ for $\isotope{Ca}$) relative
  to the full shell-model results, for selected
  states of the $\isotope{Ca}$, $\isotope{Ti}$, and
  $\isotope{Cr}$ isotopes, and for the FPD6 and GXPF1 interactions.
  These are root-mean-square averages over the full set
  of even-mass isotopes with $20< N<40$, taking all four
  $pf$-shell orbitals into account.  Calculations for $\isotope{Ca}$ are
  from~\cite{caprio2012:gssmca}. }
\label{tab-occ}
\begin{center}
\begin{ruledtabular}
\begin{tabular}{lllllllll}
\br
&\centre{4}{FPD6}&\centre{4}{GXPF1}\\ \ns\ns
&\crule{4}&\crule{4}\\ 
&
\multicolumn{1}{c}{$0^+_1$}&\multicolumn{1}{c}{$2^+_1$}&\multicolumn{1}{c}{$4^+_1$}&\multicolumn{1}{c}{$0^+_2$}&
\multicolumn{1}{c}{$0^+_1$}&\multicolumn{1}{c}{$2^+_1$}&\multicolumn{1}{c}{$4^+_1$}&\multicolumn{1}{c}{$0^+_2$}\\
\mr
\isotope{Ca}&
0.03&0.09&0.11&0.29&    
0.015&0.03&0.08&0.26\\
\isotope{Ti}&
0.10&0.18&0.17&0.37&
0.08&0.13&0.09&0.30\\
\isotope{Cr}&
0.30&0.34&0.36&0.27&
0.30&0.35&0.42&0.24\\
\br
\end{tabular}
\end{ruledtabular}
\raggedright
\end{center}
\end{table}
%----------------------------------------------------------------

In both the $\isotope{Ti}$ and $\isotope{Cr}$ isotopes, occupations
obtained in the generalized seniority scheme for the first $2^+$ and
$4^+$ states follow very similar patterns to those for the ground
state, and with a comparable (or marginally lower) level of accuracy,
summarized in table~\ref{tab-occ}.  Occupations for the first excited
$0^+$ state are obtained with comparable accuracy to that obtained in
the semimagic nuclei, and, in $\isotope{Cr}$, comparable to that for
the lower-lying states as well (table~\ref{tab-occ}).  This is perhaps
surprising, given the much poorer description of energies for the
excited $0^+$ state in the generalized seniority scheme
(table~\ref{tab-energy}).

The occupations of the \textit{proton} orbitals along the
$\isotope{Ti}$ and $\isotope{Cr}$ isotopic chains, by the two valence
protons for $\isotope{Ti}$ [figure~\ref{fig-occ}(g,h)] or four for
$\isotope{Cr}$ [figure~\ref{fig-occ}(i,j)], depend on the filling of the
neutron orbitals.  In the full shell-model space, these protons have a
significant probability of occupying orbitals other than the $f_{7/2}$
orbital, especially for $N\lesssim 28$.  As noted in
section~\ref{sec-methods}, the $\alpha_{p,a}$ coefficients in the proton
$S$ pair are nearly constant along an isotopic chain
[figure~\ref{fig-alpha}~(bottom)].  Therefore, the proton occupations
obtained in a simple $S$-pair condensate, shown in 
figure~\ref{fig-occ-v0}~(right),  are likewise nearly constant
along the chain, varying by only $\sim1\%$--$2\%$ for the
$\isotope{Ti}$ and $\isotope{Cr}$ isotopes.  However, for the $\isotope{Ti}$ isotopes,
the $(v_p,v_n)=(2,2)$ generalized seniority space [figure~\ref{fig-occ}(g,h)] largely accounts for
the actual variation in the proton $f_{7/2}$ orbital occupation, to
within $0.08$ nucleon throughout the shell.  The agreement is
significantly better for $28<N<40$, to within $0.007$ nucleon, with
differences from the full-space results averaging only $0.0046$
nucleons for FPD6 or $0.0039$ nucleons for GXPF1.  The nature of the
deviation is, as for the neutron orbitals, to overestimate the
occupation of the $f_{7/2}$ orbital.  For the $\isotope{Cr}$ isotopes [figure~\ref{fig-occ}(i,j)],
the deviations follow the same pattern, but with significantly larger
magnitudes.

\subsection{Electromagnetic observables}
\label{sec-results-trans}
%----------------------------------------------------------------
\begin{figure}
\begin{center}
\includegraphics*[width=\hsize]{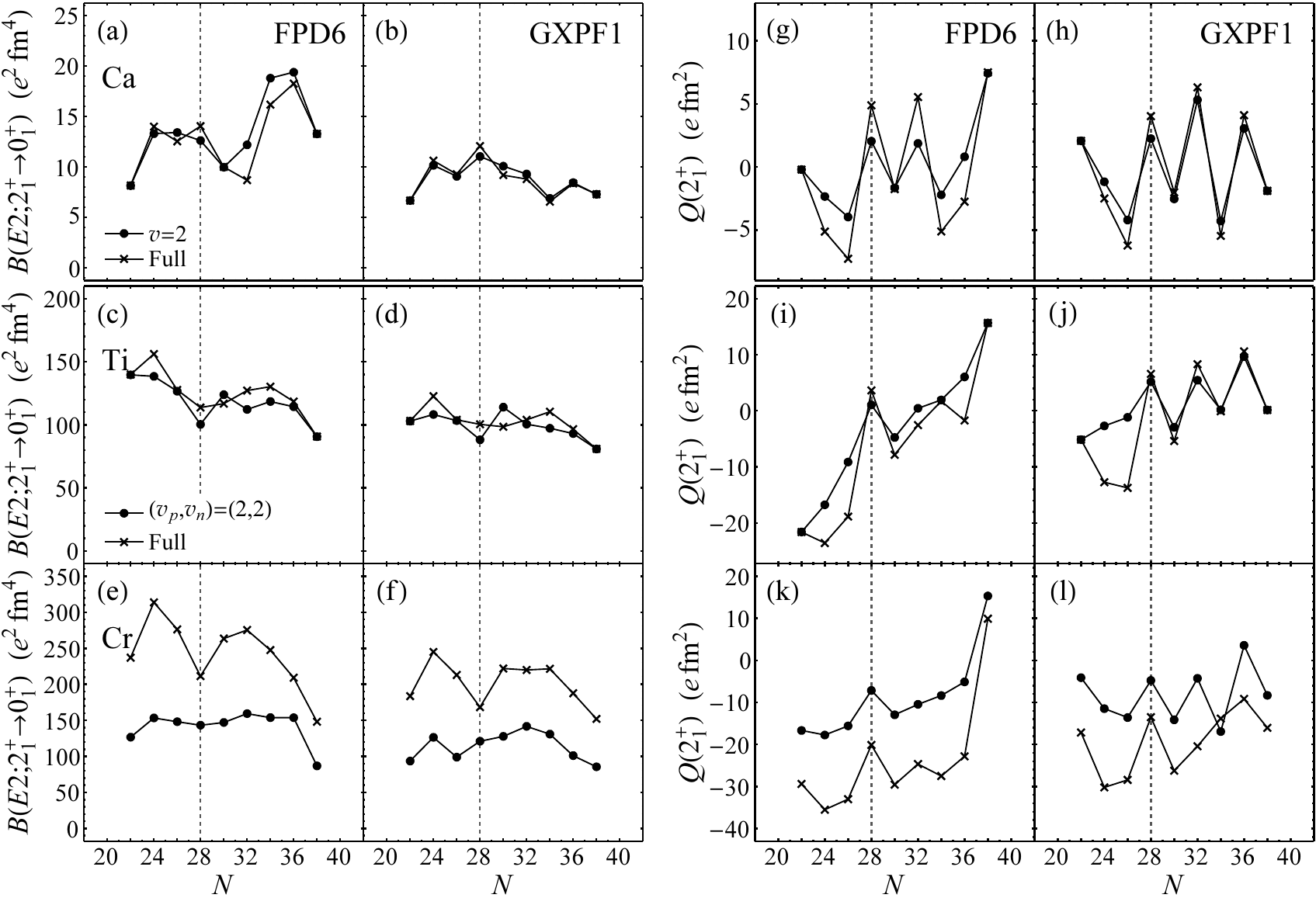}
\end{center}
\caption{
Electric quadrupole observables
$B(E2;2^+_1\rightarrow0^+_1)$~(left) and $Q(2^+_1)$~(right), calculated in the generalized seniority
$(v_p,v_n)=(2,2)$ model space (or $v=2$ for $\isotope{Ca}$) (circles) and full shell-model space (crosses).
Values are shown for the even-mass $\isotope{Ca}$, $\isotope{Ti}$, and $\isotope{Cr}$
isotopes (top to bottom, respectively), for the FPD6 and GXPF1
interactions, and are obtained using effective charges 
$e_p=1.5$ and $e_n=0.5$.
The dashed vertical line indicates the $N=28$
subshell closure. Calculations for $\isotope{Ca}$ are from~\cite{caprio2012:gssmca}. }
\label{fig-b2m2}
\end{figure}
%----------------------------------------------------------------

%----------------------------------------------------------------
\begin{figure}
\begin{center}
\includegraphics*[width=0.5\hsize]{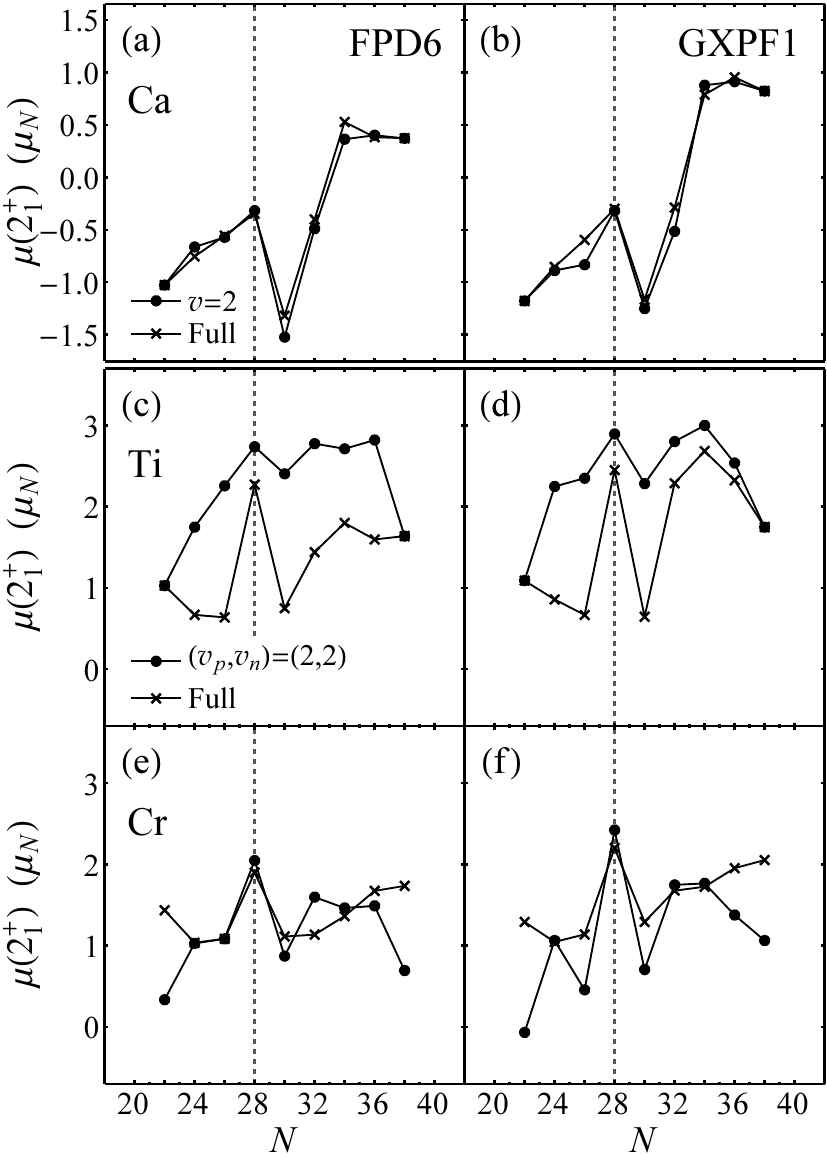}
\end{center}
\caption{
Magnetic moment 
$\mu(2^+_1)$, calculated in the generalized seniority
$(v_p,v_n)=(2,2)$ model space (or $v=2$ for $\isotope{Ca}$) (circles) and full shell-model space (crosses).
Values are shown for the even-mass $\isotope{Ca}$, $\isotope{Ti}$, and $\isotope{Cr}$
isotopes (top to bottom, respectively), for the FPD6 and GXPF1
interactions, and are obtained using free-space $g$-factors.
The dashed vertical line indicates the $N=28$
subshell closure. Calculations for $\isotope{Ca}$ are from~\cite{caprio2012:gssmca}. }
\label{fig-m1}
\end{figure}
%----------------------------------------------------------------

Electromagnetic moments and transition strengths probe correlations in
the eigenstates which are not simply apparent from the occupations of
section~\ref{sec-results-occ}.  Of particular interest is the extent to
which electric quadrupole correlations are reproduced in a space
truncated to low generalized seniority, as one moves into the shell
interior, where the proton-neutron quadrupole-quadrupole interaction
is expected to play an increasingly significant role.  
The electric quadrupole reduced transition probability
$B(E2;2^+_1\rightarrow0^+_1)$ and quadrupole moment $Q(2^+_1)$ are
shown for the $\isotope{Ca}$, $\isotope{Ti}$, and $\isotope{Cr}$
isotopes in figure~\ref{fig-b2m2}.  Electromagnetic transition matrix
elements are obtained from the one-body densities, which are evaluated
in the generalized seniority scheme as described in~\cite{luo2011:gssmme,caprio2012:gssmca}.

To review the situation for the $\isotope{Ca}$
isotopes~\cite{caprio2012:gssmca}, the evolution of
$B(E2;2^+_1\rightarrow0^+_1)$ strengths [figure~\ref{fig-b2m2}(a,b)]
across the shell is reproduced in the generalized seniority $v=2$
space in its qualitative features (minima and maxima) and with a
quantitative accuracy averaging $\sim12\%$ for FPD6 or $\sim6\%$ for
GXPF1.  The magnitude of $Q(2^+_1)$ [figure~\ref{fig-b2m2}(g,h)] is
markedly attenuated in the generalized seniority calculations relative
to the full-space results, although the generalized seniority results
demonstrate the same qualitative features as the results in the full
space, notably exhibiting the same alternations in sign as a function
of $N$.

For the $\isotope{Ti}$ isotopes, the $B(E2;2^+_1\rightarrow0^+_1)$
strengths [figure~\ref{fig-b2m2}(c,d)] are an order of magnitude larger than for the $\isotope{Ca}$
isotopes [note the change of scale from figure~\ref{fig-b2m2}(a,b) to
figure~\ref{fig-b2m2}(c,d)].  The variation with neutron number is
relatively flat, but downsloping with increasing $N$.
Both these broad characteristics are well-reproduced in the generalized seniority
$(v_p,v_n)=(2,2)$
space, although the generalized seniority calculations do not reproduce the detailed
fluctuations with neutron number.  The deviations under FPD6 are
$\sim8\%$ [averaging $10\,e^2\mathrm{fm}^4$, on 
$B(E2)$ values averaging $126\,e^2\mathrm{fm}^4$] or under GXPF1 $\sim9\%$ [averaging $9.4\,e^2\mathrm{fm}^4$, on 
$B(E2)$ values averaging $103\,e^2\mathrm{fm}^4$].
In contrast, the quadrupole moments [figure~\ref{fig-b2m2}(i,j)]
are poorly reproduced in the $(v_p,v_n)=(2,2)$ calculations, even
qualitatively, \textit{below}
the $N=28$ subshell closure.  They are somewhat better reproduced
\textit{above} the subshell closure, in particular for the GXPF1
interaction, where the detailed variation with $N$ is reproduced.  
The deviations in this range $28< N< 40$ are
$\sim50\%$ for FPD6 [averaging $4.0\,e^2\mathrm{fm}^4$, on 
$B(E2)$ values averaging $8.0\,e^2\mathrm{fm}^4$] but a somewhat smaller $\sim27\%$
for GXPF1 [averaging $1.7\,e^2\mathrm{fm}^4$, on 
$B(E2)$ values averaging $6.5\,e^2\mathrm{fm}^4$].

The $B(E2;2^+_1\rightarrow0^+_1)$ strengths, as obtained in the full
shell-model space, approximately double moving from the $\isotope{Ti}$
isotopes to the $\isotope{Cr}$ isotopes [figure~\ref{fig-b2m2}(e,f)].
However, the $B(E2)$ strengths in the generalized seniority
calculations appear to have saturated, with minimal increase relative
to the $\isotope{Ti}$ isotopes.  The magnitudes of for the quadrupole
moments [figure~\ref{fig-b2m2}(k,l)] for the $\isotope{Cr}$ isotopes are
similarly underestimated by a factor of approximately two in the
$(v_p,v_n)=(2,2)$ space, although the $N$-dependence is more complicated.  Thus, considering only one broken pair of
each type fails to produce meaningful results for electric quadrupole
observables for the $\isotope{Cr}$ isotopes.

The magnetic dipole moments $\mu(2^+_1)$, shown in
figure~\ref{fig-m1}, are
well-reproduced for the $\isotope{Ca}$ isotopes in the generalized
seniority $v=2$ space [figure~\ref{fig-m1}(a,b)].  However, for the
$\isotope{Ti}$ [figure~\ref{fig-m1}(c,d)] and $\isotope{Cr}$
[figure~\ref{fig-m1}(e,f)] isotopes, the dipole moment calculated
in the $(v_p,v_n)=(2,2)$ space has an $N$-dependence which bears little
qualitative resemblance to that of the full shell-model results.  The
best reproduction is right at the neutron $N=28$ subshell closure and,
as found above for other observables, in the upper $pf$-shell for the
$\isotope{Ti}$ isotopes with the GXPF1 interaction.  

\section{Conclusions}
\label{sec-concl}

The success of calculations truncated at low generalized seniority,
here involving one broken proton pair and one broken neutron pair,
depends on several factors.  As anticipated, the $pn$ interaction
precipitates a breakdown of the viability of such a severe truncation
with respect to generalized seniority, which is found to retain some
applicability two protons away from closed-shell ($\isotope{Ti}$) but
not four protons away from closed-shell ($\isotope{Cr}$).  Comparing
results obtained with FPD6 and GXPF1, it is seen that the particular
interaction plays a significant role.  It would be valuable to have
a systematic understanding of how this dependence might reflect
underlying quantitative properties of the
interaction.  As noted in~\cite{caprio2012:gssmca}, a
decomposition of the interaction into an appropriate set of pairing and nonpairing
(\textit{e.g.}, quadrupole) components through spectral distribution
theory, as in~\cite{sviratcheva2006:realistic-symmetries}, could yield relevant
measures.  The relative
scales of the single-particle energy differences
and the two-body interaction strengths may also play a
role~\cite{lei2011:gssm-random}.  The presence of a subshell
closure also has significant implications for the success of the
generalized seniority approximation, as previously proposed by
Monnoye \textit{et al.}~\cite{monnoye2002:gssm-ni}.  Between subshell
closures, the {filling order} of orbitals within the shell seems to be
influential, \textit{i.e.}, whether there is effectively a single
active $j$-shell, as for $20\leq N \leq 28$, or several $j$-shells are
being filled simultaneously, as in the upper $pf$ shell.  The low
generalized seniority approximations are found to miss the most
binding energy, in the present examples, for nuclei with $N\approx Z$,
\textit{e.g.}, $\isotope[48]{Cr}$, suggesting alternatively that
missing $T=0$ pairing correlations may be responsible for some of the
deviations observed in the $f_{7/2}$ nuclei.  These
correlations might be more fully accomodated in the isospin
generalized seniority scheme~\cite{talmi2001:gssm-isospin} than in the
proton-neutron scheme.

The present studies serve as a baseline for a possible improved
generalized seniority description, not simply through the inclusion of
more broken pairs, but, more fundamentally, through selection of the
collective proton and neutron $S$ pairs.  For an ideal generalized seniority
\textit{conserving} interaction in semimagic nuclei, as considered by
Talmi~\cite{talmi1971:shell-seniority}, the pair amplitudes would be
determined statically, by diagonalization in the two-particle space, and
would be constant across the shell.  The variational
prescription~\cite{gambhir1969:bpm,allaart1988:bpm} used here to
determine the pair amplitudes is a dynamical prescription, in
that it considers the effects of $pp$ interactions in determining the
proton amplitudes and $nn$ interactions in the determining the neutron
amplitudes.  However, 
the deviations in calculated occupancies, as noted in
section~\ref{sec-results-occ}, are systematic and smoothly-varying.  This
suggests that it may be possible to determine the $S$-pair amplitudes~---
which in turn determine the occupancies in the $S$-pair condensate and thus
the starting point for improvement with broken pairs~---  more optimally.  For instance, an alternative prescription for
obtaining the collective pair amplitudes by angular-momentum
projection of the Hartree-Fock-Bogoliubov intrinsic
state~\cite{maglione1983:ibm-micro,pittel1983:pair-hfb} directly takes
into account the influence of $pn$ interactions and yields
significantly modified spectra for transitional nuclei in a
nucleon pair
approximation~\cite{lei2012:npsm-transitional-PREPRINT}.
%----------------------------------------------------------------
\begin{figure}
\begin{center}
\includegraphics*[width=0.5\hsize]{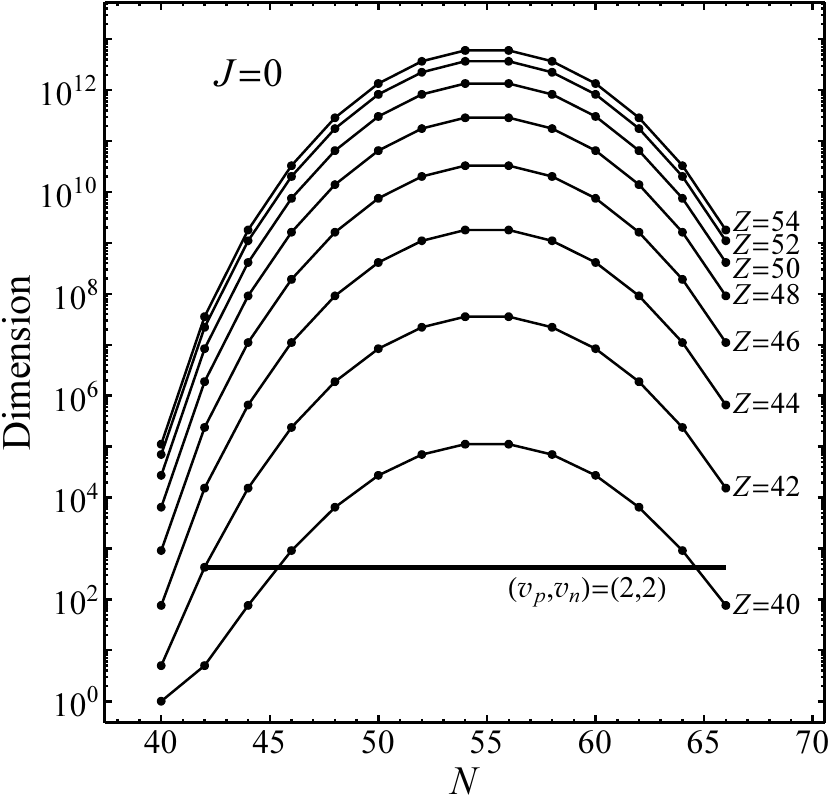}
\end{center}
\caption{
Dimensions of the $J=0$ shell-model spaces for nuclei treated in the
$sdg$ major shell.  The dimension of the generalized seniority
$(v_p,v_n)=(2,2)$ model space is shown for comparison.  Dimensions for
nuclei with $55< N,Z \leq 70$ follow from those with $40\leq N,Z <
55$ by particle-hole conjugation symmetry.
}
\label{fig-dim-sdg}
\end{figure}
%----------------------------------------------------------------

As already noted, the interest of the generalized seniority
description lies both in its value as a simple description of the
dominant BCS-like pairing correlations in a number-conserving
framework and in its potential role as a foundation for bosonized
descriptions of collective dynamics.  However, setting these
conceptual matters aside, it is also interesting to consider the
shell-model dimensions involved in the cases for which the low
generalized-seniority truncation would have the most practical utility
directly, as a computational scheme.  The present results are consistent with
the expectation that this would be for a large single-particle space
but sufficiently near a shell or subshell closure to restrict $pn$
correlations and deformation, as in the $\isotope{Ca}$ and
$\isotope{Ti}$ isotopes.  The $\isotope{Sn}$ isotopes are the
classic phenomenological example for a generalized seniority
description~\cite{talmi1971:shell-seniority}.  As a concrete numerical
example, consider the description of the light $\isotope{Sn}$ isotopes ($Z=50$), or adjacent
$\isotope{Cd}$ ($Z=48$) and $\isotope{Te}$ ($Z=52$) isotopes, in the
full $sdg$ major shell.  The dimensions for the full shell-model spaces
for various isotopic chains, including these, are shown in
figure~\ref{fig-dim-sdg}, for $J=0$.  In comparison, the
generalized seniority $(v_p,v_n)=(2,2)$ space has the same size as the shell-model
space for $\isotope[84]{Mo}$ in these same orbitals~--- 
dimension $427$ for $J=0$, $1770$ for $J=2$, $2306$ for $J=4$,
\textit{etc.}  The reduction in going from the full space to the
generalized seniority $(v_p,v_n)=(2,2)$ space reaches ten orders of
magnitude for $\isotope[104\text{--}106]{Sn}$.

%%\clearpage

%%\begin{acknowledgments}
\ack
We thank F~Iachello, S~Pittel, J~P~Vary, S~Frauendorf, and P~Van~Isacker for valuable discussions and
M~Horoi and B~A~Brown for generous assistance with \textsc{NuShellX}.  
Dimensions for figures~\ref{fig-dim-pf} and~\ref{fig-dim-sdg} were calculated using \textsc{su3shell}~\cite{dytrych:su3shell}.
This work was supported by the Research Corporation for Science
Advancement under a Cottrell Scholar Award, by the US Department of
Energy under Grant No.~DE-FG02-95ER-40934, and by a charg\'e de
recherche honorifique from the Fonds de la Recherche Scientifique
(Belgium).  Computational resources were provided by the University of
Notre Dame Center for Research Computing.
%%\end{acknowledgments}

%***************************************************************************
% bibliography
%***************************************************************************

\section*{References}

\providecommand{\newblock}{}

%bibliography{master,mc,theory,expt,books,proc,misc,gssmpf}

\end{document}